\documentclass[10pt,conference]{IEEEtran}
\usepackage{amsmath,amsfonts}
\usepackage{algorithmic}
\usepackage{graphicx}
\usepackage{textcomp}
\usepackage[usenames,dvipsnames]{xcolor}
\usepackage[hidelinks]{hyperref}
\usepackage[capitalize]{cleveref} 
\Crefname{section}{Sec.}{Sects.}
\usepackage{xurl}
\usepackage{enumitem}
\setitemize{noitemsep,topsep=0pt,parsep=0pt,partopsep=0pt}
\usepackage{todonotes}
\usepackage{booktabs}
\usepackage{multirow}
\usepackage{tabularx}
\usepackage{relsize}
\usepackage{subcaption}
\usepackage{threeparttable}
\usepackage{balance}
\usepackage{rotating}
\usepackage{microtype}

\def\BibTeX{{\rm B\kern-.05em{\sc i\kern-.025em b}\kern-.08em
    T\kern-.1667em\lower.7ex\hbox{E}\kern-.125emX}}

\newcommand\parhead[1]{\vspace{+.02cm}\noindent\textbf{{#1.}}}
\definecolor{blue(ncs)}{rgb}{0.0, 0.53, 0.74}

\newcounter{resq}
\newcommand{\resq}[1]{\refstepcounter{resq}\label[resq]{#1}}
\crefname{resq}{RQ}{RQs}
\crefformat{resq}{RQ#2#1#3}
\Crefname{resq}{RQ}{RQs}

\newcommand{\summary}[2]{
	\noindent
	\hspace{-0.185cm}
	\begin{tikzpicture}
		\node[align=left,draw,thin,minimum width=\columnwidth,inner sep=1mm] (titlebox)%
			{\parbox{\dimexpr \linewidth-2\fboxsep-2\fboxrule}{\vspace{0.15cm}\noindent\textit{#2}}};
		\node[label=right:{\colorbox{white}{\small \textbf{#1}}}] (W) at (titlebox.north west) {};%
	\end{tikzpicture}
}

\newcommand{\challengename}{Challenge} 
\newcounter{challenge}
\crefformat{challenge}{#2\challengename\ #1#3}
\Crefformat{challenge}{#2\challengename\ #1#3}

\begin{document}

\title{
	\vspace{-1cm}
	 \begingroup
	 \normalsize 
	 Accepted at International Conference on Software Engineering (ICSE) 2025 \\[1em] 
	 \endgroup
	A Large-Scale Study of Model Integration\\in ML-Enabled Software Systems
}

\author{\IEEEauthorblockN{Yorick Sens\textsuperscript{*}, Henriette Knopp\textsuperscript{*}, Sven Peldszus\textsuperscript{*}, Thorsten Berger\textsuperscript{*,$\dagger$}}
\IEEEauthorblockA{
		\textsuperscript{*}\textit{Ruhr University Bochum}, Germany \hspace{00.5cm}
		\textsuperscript{$\dagger$}\textit{Chalmers $|$ University of Gothenburg}, Sweden
}}

\maketitle

\begin{abstract}\looseness=-1
The rise of machine learning (ML) and its integration into software systems has drastically changed development practices.
While software engineering traditionally focused on manually created code artifacts with dedicated processes and architectures, 
ML-enabled systems require additional data-science methods and tools to create ML artifacts---especially ML models and training data. However, integrating models into systems, and managing the many different artifacts involved, is far from trivial. ML-enabled systems can easily have multiple ML models that interact with each other and with traditional code in intricate ways.
Unfortunately, while challenges and practices of building ML-enabled systems have been studied, little is known about the characteristics of real-world ML-enabled systems
beyond isolated examples. 
Improving engineering processes and architectures for ML-enabled systems requires improving the empirical understanding of these systems.

\looseness=-1
We present a large-scale study of 2,928 open-source ML-enabled software systems. We classified and analyzed them to determine system characteristics, model and code reuse practices, and architectural aspects of integrating ML models. Our findings show that these systems still mainly consist of traditional source code, and that ML model reuse through code duplication or pre-trained models is common. We also identified different ML integration patterns and related implementation practices. We hope that our results help improve practices for integrating ML models, bringing data science and software engineering closer together.
\end{abstract}


\section{Introduction}
\label{sec:introduction}
\noindent
\looseness=-1
Many recent breakthroughs in machine learning (ML) have given rise to ML-enabled software that was not realizable before.
Consider cyber-physical systems, such as autonomous vehicles\,\cite{chernikova2019self,gupta2021deep} or unmanned aerial vehicles\,\cite{chowdhuryengineering}, as well as software in finance\,\cite{vellido2020importance} or healthcare\,\cite{goodell2021artificial}.
All these systems benefit from advances in ML model architectures and training algorithms.
Nevertheless, developers still struggle when de\-velo\-ping software systems that integrate ML models\,\cite{ryseff2024failure,kahn2022}.
In fact, industrial surveys report that 78\,\% of ML projects stall\,\cite{News2019}, while others say as many as 85\,\% fail\,\cite{ONeill2019}.
Many factors can cause such failures\,\cite{Hotz2024}, including development processes\,\cite{amershi2019software, becker2017predicting, arpteg2018software}, data quality\,\cite{bosch2021engineering}, system architecture\,\cite{gorton2023software}, tool support\,\cite{idowu2022effectiveness,idowu2022asset,zhao2024empirical,alves2023practices,sculley2015,bhatia2022towards}, quality assurance\,\cite{braiek2020testing}, and the integration of ML models with traditional software components\,\cite{nazir2024architecting, nahar2023meta}---the focus of our work.



\looseness=-1
Integrating ML models in software systems is challenging. Executing the models requires boilerplate code and extra in\-fra\-structure\,\cite{sculley2015}. The lack of behavioral guarantees in ML models---being probabilistic by nature---requires additional safeguards, especially in safety-critical systems\,\cite{braiek2020testing}. Safeguards are typically implemented manually to prevent undefined behavior for certain inputs, as well as invalid results that would violate domain restrictions. 
These challenges are further exacerbated by systems that integrate multiple models. 
Extreme examples are perception systems in autonomous driving, some of which boast up to 28 different ML models that interact with each other\,\cite{peng2020apollo} and are safeguarded with manually implemented heuristics.
Furthermore, training and developing ML models is often costly\,\cite{hendrycks2019using, erhan2010does}, which made ML model reuse a common practice, 
In particular, reusing pre-trained ML models allows leveraging larger models trained on more data, but does not fit well with current software engineering practices. 
Unfortunately, empirical insights on reuse practices are missing.

\looseness=-1
Research on engineering ML-enabled systems has focused on challenges and practices of engineering such systems\,\cite{amershi2019software,bosch2021engineering,
braiek2020testing,arpteg2018software,zhao2024empirical,nazir2024architecting}. Researchers also created novel workflows and tooling (e.g., experiment management tools\,\cite{idowu2022asset,idowu2022emmm}) to address the challenges. Much work also focused on dependability and safety of ML models\,\cite{martinez2022software}. 
However, building ML-enabled systems requires holistic methods and tools---to systematically integrating traditional software and data-science artifacts in a whole system, as well as maintaining and evolving such systems.
Progress in this area has been limited so far, which can be attributed to a lack of understanding of the current practices and concrete problems of developing ML-enabled software systems. In other words, while researchers have focused on ML engineering, still, little is known about how ML models are integrated into ML-enabled systems in practice.
%
%

We present an empirical study of 2,928 ML-enabled software repositories on GitHub, characterizing the use of ML in software systems. We focused on the reuse of ML models across software systems, as well as the integration of ML models. Our analysis addresses three main research questions:

\textbf{\textit{RQ1}:}\resq{it:rq1} \textit{What are characteristics of ML-enabled software systems?}
To understand such systems, we studied what role ML plays in them and what characterizes them.
For instance, what is the proportion of ML code in relation to non-ML source code,
what types of software systems employ ML (e.g., libraries or end-user-oriented applications), and how?

\textbf{\textit{RQ2}:}\resq{it:rq2} \textit{How are ML models reused across software systems?}
The growing size (e.g., number of trainable parameters) of ML models both enables and necessitates their reuse across software systems---as they become more general in their capabilities, but also more difficult to train\,\cite{qiu2020pretrainednlp,long2022pretrained}. We explored how reuse is done in practice, like reuse of source code has been studied in previous works\,\cite{frakes2005reuse, mojica2013large}.

\textbf{\textit{RQ3}:}\resq{it:rq3} \textit{How is the integration of ML models reflected in architectural aspects of software systems?}
We explored how ML models are integrated into systems and what functionality they realize. We identified how many models are used, how they interact, and how they are integrated with code, for instance, with pre- and post-processing methods.

\looseness=-1 We contribute:
\begin{itemize}
	\item insights on the \textbf{share of ML-related code} of ML-enabled systems and its distribution, showing that even \textbf{in ML-enabled systems traditional source code makes up for most parts of the system};
\item insights into \textbf{ML reuse practices}, highlighting a \textbf{strong reliance on pre-trained models}, as well as the practice of \textbf{copying source code implementing ML models} between systems;
\item a \textbf{catalog of ML integration patterns} and coding practices derived from an \textbf{in-depth analysis of 26 applications} that use ML to provide end-user-oriented functionality;
\item and a \textbf{dataset of 2,928 ML-enabled software systems}, of which we manually classified 160 according to the type of the system and their relation to~ML.
\end{itemize}

For replication and further studies, we provide a replication package, including the dataset and analysis scripts\,\cite{replication}.
We hope that our findings will help practitioners and researchers better understand the integration of ML models in ML-enabled systems, and that our data will be useful for further research.

\section{Motivation and Research Questions}
\noindent
We present the necessary background, the state of research, as well as motivate and refine our research questions.

\looseness=-1
ML-enabled systems contain ML-based func\-tionality, ranging from utility functions to application logic.
This functionality can include predictions, data analysis, or generative content creation\,\cite{feuerriegel2024generative}. ML-enabled systems incorporate various ML technologies, such as Deep Learning (DL), Convolutional Neural Networks (CNNs), or Transformers\,\cite{lecun2015deep,yamashita2018convolutional,vaswani2017attention}.
This integration of ML into software \textit{poses new challenges for development processes, architectures, and testing, among others}\,\cite{bosch2021engineering}.
Many originate from the fundamentally different nature of ML models compared to traditional software\,\cite{nahar2023meta}.
ML models are probabilistic, essentially constituting unreliable functions, while traditional software is more deterministic.

\label{sec:background}
\looseness=-1
\parhead{Characteristics of ML-Enabled Software Systems (RQ1)}
While there exists extensive theory in data science, the majority of techniques is model-centric. However, integrating ML models into software systems to provide value to end-users requires system-centric engineering methods.
Effective methods need to be tailored towards the actual characteristics and properties of ML in the context of systems.
To this end, we need to improve our empirical understanding of ML-enabled systems in general.
	Furthermore, software systems can be of different types, such as applications, libraries, and frameworks \cite{idowu2024large, nahar2024product}, which impacts the use of ML.
\textit{\textbf{RQ1.1} (What types of systems use ML and how are they related to it?)} first determines what systems\,\cite{idowu2024large,nahar2024product} (e.g., applications or libraries) integrate ML models and what ML is used for (e.g., for end-user functionality or purely conceptual contributions). This puts the results of our study into context and helps identify differences in the use of ML, as well as implications for practice.

\looseness=-1
ML-enabled systems rely on ML libraries and frameworks. The most popular ones\,\cite{stancin2019libraries} are PyTorch\,\cite{pytorch}, TensorFlow\,\cite{tensorflow}, and Scikit-Learn\,\cite{scilearn}. Many others build upon them. While TensorFlow and PyTorch offer DL, Scikit-Learn offers different techniques, mostly traditional ML ranging from supervised methods (e.g., linear regression and perceptron) to unsupervised methods (e.g., clustering and PCA).
TensorFlow and PyTorch are frameworks designed for a modular implementation of ML models. A class ``Module'' (\emph{``BaseEstimator''} in Scikit-Learn) represents either complete models or their building blocks (e.g., layers).
Developers typically extend this class, often using predefined modules, such as linear or convolution layers.

\looseness=-1
ML-enabled systems contain additional artifacts (called \emph{ML assets}\,\cite{idowu2022asset} in the remainder),
including model implementations, model binaries, ML training code, and other ML files.
These ML assets must be integrated with code and co-evolved.
\textit{\textbf{RQ1.2} (How much of ML-enabled systems is specific to ML?)} elicits details on the relevance, type, and characteristics of ML assets. 

\looseness=-1
Finally, while quality assurance is an integral part of ML model training, also the system has to be considered.
However, the concrete quality assurance practices in ML-enabled systems
have not been captured on a large scale, beyond individual qualitative studies\,\cite{openja2024empirical}.
To obtain insights on these practices, \textit{\textbf{RQ1.3} (What are quality assurance practices of ML-enabled systems?)} investigates quality assurance on a large scale.

\looseness=-1
\parhead{Reuse of ML Models (RQ2)}
Like reuse in traditional software\,\cite{mojica2013large}, developers reuse ML assets.
%
\looseness=-1
It is commonly known that there is reuse of established ML models, allowing developers to use them without the extensive effort required to build a new model\,\cite{long2022pretrained,qiu2020pretrainednlp}. Such models are called pre-trained models, as they are optimized and fully functional.
Pre-trained models for specific tasks are distributed through model hubs. Jiang et al.\,\cite{jiang2022reusehub} identified 8 model hubs when studying the content of their provided artifacts and related security risks: PyTorch Hub\,\cite{torchhub}, Tensorflow Hub\,\cite{tensorflow}, Hugging Face\,\cite{huggingface}, Model Zoo\,\cite{modelzoo}, ONNX Model Zoo\,\cite{onnxmodelzoo}, Model Hub\,\cite{modelhub}, NVIDIA NGC\,\cite{nvidia}, and MATLAB Model Hub\,\cite{matlab}.
While we know all these sources from which pre-trained models could be reused, we know little about the actual practices, neither how prevalent this reuse is nor how it is realized.
\textit{\textbf{RQ2.1} (To what extent are pre-trained models used?)} addresses this gap.

Apart from full models, other ML assets (e.g., scripts, model binaries) can be reused as well.
However, there are currently no studies on such reuse and its relation to model reuse, e.g., reusing implementations of ML models as an alternative to pre-trained models.
Related studies focus on how to identify model plagiarism\,\cite{li2021modeldiff}, or on reuse practices of models from specific libraries or platforms\,\cite{jiang2023huggingface}.
Consequently, we need to improve our empirical understanding to what degree ML models are currently reused, what other assets are also reused, and how that reuse is realized---addressed by \textit{\textbf{RQ2.2} (To what extent are ML assets reused between ML-enabled systems?)}. 

\looseness=-1
\parhead{Architectural Aspects (RQ3)}
The architecture of ML-enabled systems revolves around ML models, ML assets, and traditional software, which must be integrated together.
However, little is known on how the interaction of ML models can impact the behavior of a system.
While some architectures for ML-enabled systems have been proposed, they usually do not focus on concrete interactions and topologies of ML models, but rather on high-level system abstractions.
This includes, for instance, a microservice architecture, where ML models are modularized and interact with the rest of the system via REST\,\cite{ribeiro2019microservicearch}.
Others suggest separating ML logic from the business logic of a system\,\cite{yokoyama2019mlarchitecture}.

ML models are typically encapsulated in software modules that contain the ML model (binary) along with source code for model loading and execution, including required pre- and post-processing steps.
Common anti-patterns include excessive glue code, complicated data-processing pipelines, dead experimental code paths, and lack of abstraction\,\cite{sculley2015}.
Since ML models are updated or replaced regularly\,\cite{Ozkaya2020}, i.e., by an optimized version, which in turn affects the remaining system (e.g., thresholds in glue code need to be adjusted), such anti-patterns hinder optimization.
Little is known about best practices in such situations.
To derive such, \textit{\textbf{RQ3.1} (How are ML models embedded into traditional code?)} investigates current ML model integration patterns.

Complex systems can integrate many ML models. Recall perception systems with up to 28 models, integrated in complex topologies, with non-trivial interactions and dependencies between models\,\cite{peng2020apollo}.
While current research discusses the implications of such topologies (e.g., propagating error, computational complexity)\,\cite{Peldszus2023, apel2022feature}, it is important to study these in a larger dataset and define common patterns---addressed by \textit{\textbf{RQ3.2} (What are interaction patterns of ML models?)}.

\section{Methodology}
\label{sec:method}
\begin{figure}[t]
	\centering
	\includegraphics[width=\linewidth]{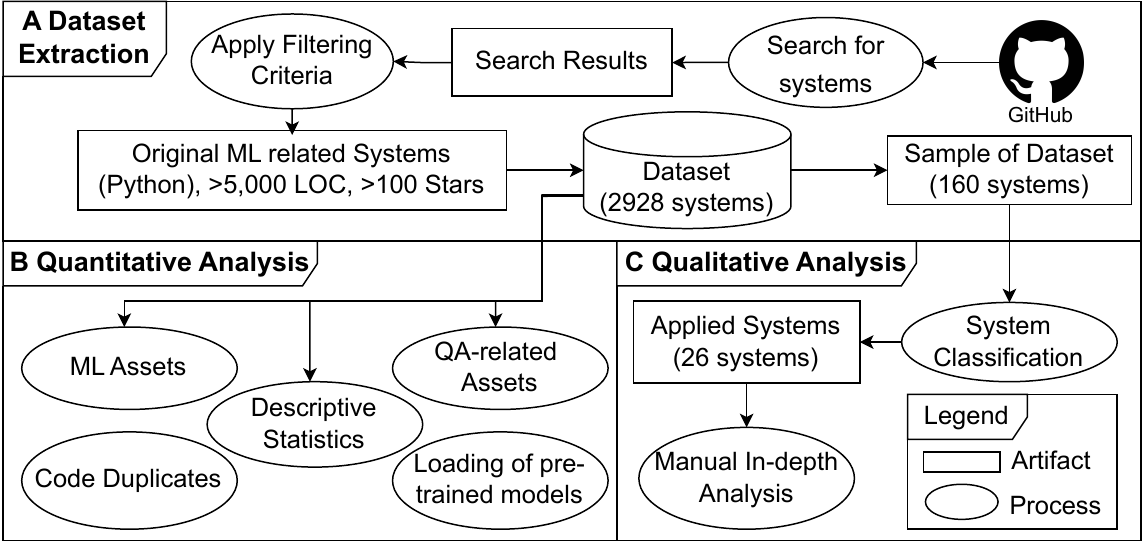}
	\vspace{-12pt}
	\caption{Methodology overview}
	\label{fig:methods}
	\vspace{-12pt}
\end{figure}

\noindent
Our methodology, summarized in \cref{fig:methods}, comprised the mining of repositories with ML-enabled systems from GitHub,  and their quantitative and qualitative analysis.
While some sub-RQs could be answered by analyzing the whole dataset automatically, other sub-RQs required qualitative analysis of random samples.

\subsection{Subject Selection}
\label{subsec:acquisition}
\noindent
We mined repositories from GitHub as follows, considering typical mining strategies\,\cite{kalliamvakou2014promise}, \cite{dabic2021}.

\parhead{Inclusion Criteria}
We used GitHub's dependency graph to obtain repositories that use one of the three most popular ML libraries\,\cite{stancin2019libraries} (\labelcref{it:ic1}) and that are implemented in the currently most popular programming language for ML (Python) (\labelcref{it:ic2}).

\begin{enumerate}[label*= IC{\arabic*}, leftmargin=*]
	\item \textit{The repository depends on TensorFlow \cite{tensorflow}, PyTorch \cite{pytorch}, or scikit-learn \cite{scilearn}.} This criterion aimed to identify systems implementing or using ML technologies. ML libraries support developing (especially training) ML models, but are also needed to execute the models. \label{it:ic1}

	\item \textit{The main programming language is Python.}\label{it:ic2} This criteria aimed to ensure the relevance of our analysis---since Python is the currently most popular language for ML-enabled systems\,\cite{raschka2020machine,gonzalez2020state}---and the comparability of the subject systems by focusing only on one language.
\end{enumerate}

\noindent
Applying \labelcref{it:ic1} yielded 754,161 repositories, of which 255,096 dependent on PyTorch, 233,779 on TensorFlow, and 465,980 on Scikit-Learn, and \labelcref{it:ic2} in 405,091 Python repositories.

\parhead{Exclusion Criteria}
We then filtered the obtained repositories to create a high-quality set of ML-enabled software systems of a certain size and maturity.
We filtered out repositories if at least one of the following criteria was met.
\begin{enumerate}[label*= EC{\arabic*}, leftmargin=*]
	\item \looseness=-1\textit{The repository is not original, i.e., forked or duplicated from another repository in the dataset.\label{it:ec1}} Forks would distort the results, as popular projects with many forks would be counted multiple times and disproportionately affect the results, without providing additional insights.
	\item \textit{The repository has fewer than 100 stars.}\label{it:ec2} Sufficient popularity ensures to avoid practically irrelevant projects.
	\item \textit{The codebase contains fewer than 5,000 lines of Python code.} \label{it:ec3} This criterion aimed at ensuring a certain size of the code, excluding tutorials and toy projects.
\end{enumerate}

\looseness=-1
The thresholds for \labelcref{it:ec2} and \ref{it:ec3} were chosen based on previous studies\,\cite{gonzalez2020state, aghili2022studying, nahar2024product, biswas2021art, borges2018star} and manual exploration of a sample. We set them relatively low to exclude simple toy projects, but still obtain a large, representative dataset.

\looseness=-1
We excluded forks with \labelcref{it:ec1} in two steps: First, we removed projects that were forked directly from GitHub, which can be retrieved through GitHub's REST API.
Second, since we found that the dataset still contained a significant number of duplicates, we identified them based on identical README files and kept only the repository that was created first.
This selection resulted in a final dataset of 2,928 repositories.

\subsection{Quantitative Analysis}
\label{subsec:methods_characteristics}
\noindent
We analyzed the ML-enabled systems by creating analysis scripts and using automated tools. That was possible for the following research questions, which we answered as follows.

\looseness=-1
\parhead{Descriptive Statistics} To put all our results into context---and to determine to what kinds of systems they apply---we first created descriptive statistics.
\label{subsubsec:statistics}
Typical descriptive statistics are scale and popularity of the subjects\,\cite{vidoni2022systematic}. For popularity, we used GitHub Stars, which can also be seen as an indicator of maturity. To determine the systems' scales, we used common code metrics\,\cite{vidoni2022systematic}:
number of commits, source files, lines of code, number of classes and functions.
These metrics help to contextualize results and highlight the scale and relevance of our dataset.
We also provide statistics on the use of ML libraries, also whether multiple libraries are used.

\looseness=-1
\parhead{General Characteristics (\textit{RQ1})} The first research question, that can be answered quantitatively on the full dataset, is
\textit{\textbf{RQ1.2} (How much of ML-enabled systems is specific to ML?)}, since we were able to automatically identify ML assets, use metrics to measure their characteristics, and compare those to traditional software assets.
To identify the ML-related parts of our subject systems, we extracted ML assets by searching for the following typical ML assets:

\begin{description}
	\item[\textit{Implementations of ML models:}] Recall that models are implemented by extending the class \emph{``Module''} (PyTorch and TensorFlow) or \emph{``BaseEstimator''} (Scikit-Learn) (\cref{sec:background}).
	We searched for class definitions that inherit from these base classes or any of their subclasses provided in the libraries or defined in the subject systems.
	\item[{\textit{ML-related functions:}}] We extracted all functions and class methods that either use an API of an ML library or instantiate an ML model. Functions defined within an ML model are considered ML-related, regardless of whether the specific function directly interacts with an ML library.
	\item[{\textit{Stored model files:}}] ML models are often stored in binary files for later reuse, typically using the extensions .pt or .pth (PyTorch) and .h5 or .pb. (TensorFlow). We counted the number of such files present in the systems.
\end{description}

\looseness=-1\noindent
We compared the general code metrics of the traditional code assets to those of the identified ML assets. This analysis provides insights into the extent of the ML-specific assets compared to traditional software assets.

We also answered \textit{\textbf{RQ1.3} (What are quality assurance practices of ML-enabled systems?)} quantitatively on our full dataset, using automated techniques.
\looseness=-1
We analyzed to what degree the source code, especially the ML-related part, is covered by unit tests, and how the models are evaluated.
To identify test cases, we searched the source code for implementations of Python's unittest framework.
We then identified the unit under test by analyzing the method calls invoked.
This allowed us to measure the number of functions directly covered by tests for ML-and non-ML functions, respectively.
To investigate quality assurance of ML models, we extracted usages of ML validation functions, following another work analyzing ML-related projects\,\cite{idowu2024large}.

\looseness=-1
\parhead{Reuse of ML Models (\textit{RQ2})} The research question \textit{\textbf{RQ2.1} (To what extent are pre-trained models used?)} was answered quantitatively on the full dataset, since we were able to detect the presence of pre-trained models automatically.
Recall that using pre-trained models distributed through model hubs (\cref{sec:background}) is a systematic way of model reuse.
We relied on the list of Jiang et al.\,\cite{jiang2022reusehub}.
From each hub's documentation we extract a list of APIs used to load pre-trained models in Python. The full list can be found in our replication package\,\cite{replication}.
We then identified which systems use the APIs and where the model is loaded in the system, collecting a list of all related files.
Furthermore, we analyzed the source of the API calls in the system by categorizing them as test, demonstration, and other directories, according to their name.
We report the number of systems that contain an API used to load pre-trained models and where they are loaded.

\looseness=-1
We also analyzed reuse of ML assets at a large scale among all systems in our full dataset, answering \textit{\textbf{RQ2.2:} (To what extent are ML assets duplicated between ML-enabled systems?)} quantitatively.
Specifically, to identify duplicated ML code, we applied a code clone detection tool to the ML-enabled systems.
We decided to use JPlag\,\cite{prechelt2002jplag}, which supports Python and is ranked as one of the best tools in multiple literature reviews\,\cite{zakeri2014similarity,novak2019source, Ragkhitwetsagul2018comparison}.
In a pairwise comparison, a match is found when a sequence of tokens of a certain minimum length for two files match.
We used a threshold of 100 tokens to avoid false positives.
Based on these results, we identified the most prominent sources of duplicated model implementations by sorting the subject systems by the number of duplicated tokens shared with all others systems.
Then, starting from the projects with the highest amount of duplicated code, we manually inspected each system, to determine if they are original or what system they are based on. To this end, we analyzed the copied source files and searched for hints on the original system. Usually, the copyright information was left in place, crediting the original repository.
This analysis was repeated until we encountered no more original systems for 10 times in a row.

\looseness=-1
We report the number of code duplicates from each original system in our dataset.
Since some original systems were not contained in the dataset (e.g. because they were not stored on GitHub), we use another system, that only contains a complete duplicate from this system and no other, as a placeholder to report the number of duplicates from the original one.

\subsection{Qualitative Analysis: System Classification}
\label{subsec:methods_classification}
\noindent
For our qualitative analysis, we selected a random sample of 160 systems (5.5\,\% of our dataset) and answered the following research questions requiring qualitative analysis.

To answer \textit{\textbf{RQ1.1} (What types of systems use ML and how are they related to it?)} we classify systems inspired by studies\,\cite{idowu2024large,gonzalez2020state, Rzig2022, bhatia2022towards} that investigated properties of ML-enabled systems, such as ML model development stages, and quality assurance.
More specifically, we classify according to the following two dimensions: \textit{type} of systems (i.e., Application or Library) and their \textit{relation to ML} (i.e., what are the ML based functionalities and how are they used).

\looseness=-1
We manually examined and labeled the sample in several iterations, in which the definitions of system type and relation to ML were continuously adjusted.
Two authors classified batches of systems independently and then discussed their disagreements. A third author was the mediator if no consensus was reached.
We started with 100 projects to develop an understanding of what is present in the dataset and to define the labels.
Thereafter, the two authors classified batches of 15 systems independently.
We used Krippendorff's Alpha\,\cite{McDonald2019, Krippendorff2019} for the agreement score.
As the score was below 0.6 in the first iterations, the authors discussed their disagreements.
After four iterations we achieved an alpha of 0.86 for the system type and an alpha of 0.70 for the relation of ML, confirming a common knowledge base and correctness of the assigned labels.
Based on the improved understanding, the previous systems (including the initial sample) were reclassified by one author, with the second author checking the results.

\subsection{Qualitative Analysis: Manual In-Depth Analysis}
\label{subsec:methods_manual}
\noindent\looseness=-1
To answer \textbf{\textit{RQ3}} and its sub-research questions \textit{\textbf{RQ3.1} (How are ML models embedded into traditional code?)} and \textit{\textbf{RQ3.2} (What are interaction patterns of ML models?)} a manual analysis of the source code was necessary to identify relations of ML models and traditional software components. Since analyzing 160 systems manually is not feasible we selected all systems that are end-user-oriented applications and that use ML to realize business logic (type ``Application'' and label ``Business-Focused Applied,'' cf. \cref{subsec:methods_characteristics}). For the selected systems, we collect insights on architectural patterns and report concrete details through examples.
Our analysis is inspired by Peng et al.\,\cite{peng2020apollo}, who studied various aspects of model integration in an autonomous driving perception system (cf. \cref{sec:background}). Upon the research questions and our experience, we defined relevant aspects to extract.
We then systematically examined the implementation of each system and documented observations related to the aspects (the raw data is in our appendix\,\cite{replication}).
We time-boxed the manual analysis to two hours per system.
If we could not gather enough information in time, for instance, due to insufficient documentation or poorly structured source code, we noted the respective aspect as unknown.

\looseness=-1
To structure the analysis, we first identified the code related to the ML models, as described in \cref{subsec:methods_characteristics}, as entry points for manual exploration.
For each entry we analyzed the surrounding code to understand the model integration.
Then, using the IDE's call-graph navigation, we determined where models are instantiated.
We disregarded instantiations that are tests or demonstrations, or that occur in files that are no longer used but are kept by developers. For instantiations that occur in actual production code, we then analyzed the surrounding code to determine the input and output of the model. If necessary, we followed the method calls further to clarify the data format of the input and output, as well as possible pre- and post-processing steps, characterizing the data flow and topologies of the models in a system. After checking each instantiation for each model, we summarized the results to obtain the total number of models used.
All projects are details in our online appendix\,\cite{replication}.

We collected the following aspects. The former are descriptive aspects putting the results from the quantitative analysis (\ref{subsec:methods_characteristics}) into context. The latter are related to model integration, answering \textit{\textbf{RQ3.1}}, and interaction patterns, answering \textit{\textbf{RQ3.2}}.

\parhead{Descriptive Aspects} While these aspects are related to general characteristics of ML-enabled systems (\textbf{\textit{RQ1}}) and practices on model reuse (\textbf{\textit{RQ2}}), our qualitative analysis goes into much more technical detail on the sample of 26 systems.

\textit{\textbf{ML Functionalities and Tasks:} What are the ML models used for? What functionality is implemented with ML? Based on which ML task is the functionality implemented?}

\textit{\textbf{Number and Type of Models:} How many ML models are used? What is their model architecture (CNN, RNN, etc)? How are they trained (supervised, unsupervised, etc.)?}

\textit{\textbf{Model Origins:} Is the model custom-built, completely pre-trained or pre-trained with additional fine-tuning?}
Besides quantitative statistics about model reuse (\textbf{\textit{RQ2}}), we complemented these with qualitative aspects (e.g., how tailored).

\parhead{Model Embedding} The following aspects address \textit{\textbf{RQ3.1} (How are ML models embedded into traditional code?)}

\textit{\textbf{Storing \& Loading of Models:} Are the models stored locally in the repository or loaded from a remote source, and how?}
Models need to be instantiated (e.g., load model parameters). While we investigated this aspect quantitatively with RQ2, we went into more detail and analyzed model instantiation qualitatively (e.g., what and how APIs are used), to understand their integration into traditional code.

\textit{\textbf{Model In- \& Output:} What is the input and output of ML models? What data types are used?}
How ML models are embedded into a system also depends on the data being processed.
We examine the type of input and output data to understand what data is used in ML-enabled functionality.

\looseness=-1
\textit{\textbf{Pre- \& Postprocessing:} What pre-and postprocessing steps are implemented for the ML models?}
Since different parts of a system typically use different data formats, to actually use data as input for an ML model, and to use its output, it often needs to be processed before and after.
Also, other functions that need to be performed on this data, such as safeguards\,\cite{Poth2020,Abdelkader2024}, are often considered essential.
This aspect aims to understand the specifics needed for integrating an ML model.

\parhead{Interaction Patterns} Our final aspect addresses \textit{\textbf{RQ3.2} (What are integration patterns of ML models?)}.

\textit{\textbf{Model Interaction:} How do multiple models interact?}
We specifically analyzed systems that contain multiple ML models with respect to the interaction of their models.
As described above, we followed call graphs to comprehend the code and to determine data flows between the models.
We sketched the topologies and when similar ones appeared in at least two systems defined these as a pattern.
The patterns from the case study by Peng et al.\,\cite{peng2020apollo} provided a basis.

\subsection{Threats to Validity}
\label{sec:threats}
\noindent\looseness=-1
\textbf{\textit{Internal Validity:}}
Personal biases of the authors, e.g., expectancy bias, may have affected the selection of subject systems and how these were analyzed quantitatively.
To mitigate this threat, we based selection criteria on external sources, such as studies that identify the most popular ML libraries\,\cite{stancin2019libraries}.
Similarly, we built up our analysis on previous works\,\cite{idowu2024large,peng2020apollo} and reused existing tooling, e.g., JPlag\,\cite{prechelt2002jplag}, to avoid internal bias in the analysis.
The classification of the 160 systems may be subject to author bias, potentially impacting how the systems are classified, and different perspectives of the labeling authors may impact the validity of the labels.
To address author bias, we involved multiple authors in labeling the systems and held frequent discussions on the labels, involving a third author as mediator. To address different perspectives, we only continued with independent labeling after reaching a sufficiently high Krippendorff's Alpha.
Author bias may also affect the manual analysis of the selected systems (\cref{subsec:methods_manual}), i.e., impacting what is considered relevant.
To mitigate this threat, we structured the analysis using aspects and questions, and frequently discussed our findings among three authors.

\noindent\looseness=-1
\textbf{\textit{External Validity:}}
Our quantitative analysis is limited to Python systems and three ML libraries (PyTorch, TensorFlow, and Scikit-Learn), which threatens the generalizability of the results.
However, since both Python and the selected libraries are most widely used for ML-enabled systems\,\cite{raschka2020machine,gonzalez2020state,stancin2019libraries}, our research covers the majority of relevant systems.
Due to the sample size of 160 of 2,928 systems, the observed distributions of system type and relation to ML are only representative with a relatively high margin of error (up to 7.43\,\% at the 95\,\% confidence level for \textit{Conceptually Applied}).
The manually analyzed subset of systems is so small that we can only present qualitative observations.
Our automated analysis relies on parsing Python code into ASTs, which failed for individual files, but the error rate (on average 0.06\,\% of the files)
is negligible.
Analyzing the number of ML-related functions is difficult due to the use of wrappers, e.g., one function loads a model and another one uses it.
As a result, we only show a lower bound on the number of ML functions, and more accurate qualitative observations as part of our manual analysis.
The sometimes generic names of the APIs used to load pre-trained models may lead to false positives when identifying the use of pre-trained models.
To address this, we only consider API calls in the context of a model hub and are not ambiguous about the loaded object.
\section{System Characteristics (RQ1)}
\label{sec:rq1}
\noindent
We now present characteristics of the systems, including their types, ML assets, and quality assurance practices.

\begin{figure}[t]
	\centering
	\includegraphics[width=\linewidth]{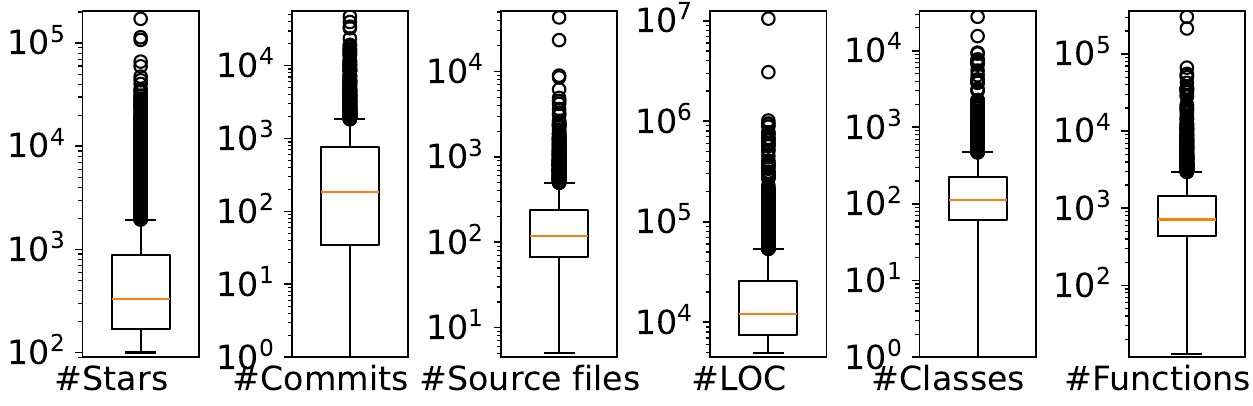}
	\vspace{-.3cm}
	\caption{General statistics of the systems in our dataset}
	\label{fig:statistics}
	\vspace{-.3cm}
\end{figure}

\begin{figure}[b]
	\vspace{-.3cm}
	\centering
	\includegraphics[width=.49\linewidth]{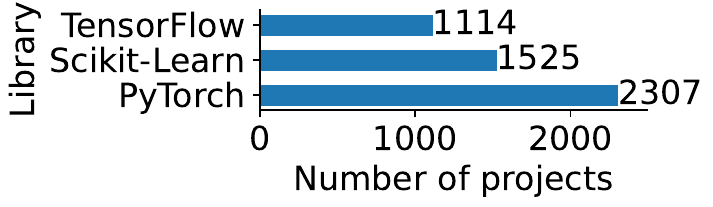}
	\hfill
	\includegraphics[width=.49\linewidth]{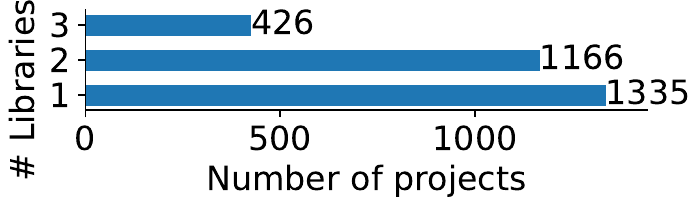}
	\vspace{-.3cm}
	\caption{ML libraries used in the ML-enabled systems}
	\label{fig:libs}
\end{figure}

\parhead{Descriptive Statistics}
\label{subsec:results_statistics}
We mined 2,928 ML-enabled systems that vary in scale significantly (see \cref{fig:statistics}).
On average, they have 263 source files (median 119), ~34,000 LOC (median ~12,000), 252 classes (median 113), and 1,171 functions (median 715).
They are of substantial popularity, with an average of 1,487 stars (median 327) and 829 commits (median 187).
Outliers are \textsf{The Algorithms} (181k stars), a collection of ML algorithms for educational purposes, and \textsf{Azure SDK for Python} with the largest codebase of 10M LOC.
\Cref{fig:libs} shows the used ML libraries.
PyTorch is most common (78\,\% of systems), followed by scikit-learn and TensorFlow (52\% / 38\% of systems).
More than half of the systems use more than one library, indicating that different types of ML are used.

\subsection{System Classification (RQ1.1)}
\label{subsec:system_types}
\noindent
Upon our random sample of 160 systems we derived labels for system types and their relation to ML.
Our appendix\,\cite{replication} has detailed label descriptions, examples, and labeling guidelines.
\Cref{tab:results_classification} shows the results of the classifications.

\parhead{System Types}
We define the following four types of systems.

\textbf{\textit{Applications}} are standalone, executable systems that provide end-user-oriented functionality through a user interface (e.g., GUI or command line).
They do not require users to have programming skills. Applications solve problems and provide solutions to end-users, implementing business logic to do so.

\looseness=-1
\textbf{\textit{Libraries}} provide functionality intended to be used in other programs via code APIs.
Except for the provided functionality they contain no application logic and cannot run standalone.

\textbf{\textit{Frameworks}} provide general application logic, but are designed to be extended by concrete functionality orchestrated by the framework. To this end, frameworks take control of the code that uses the framework, according to the principle of dependency inversion \cite{riehle2000framework}. Similar to libraries, frameworks usually provide generic helper functions that are frequently used in the context of the framework.

\textbf{\textit{Plugins}} extend applications with functionality via extension points, and only function through them. 
Users interact with plugins through the UI of the applications.

Our dataset consists mainly of libraries (62\,\%), followed by applications and frameworks (20\,\%), and almost no~plugins.
\smallskip

\parhead{Relation to ML}
We identified three ways in which an ML-enabled system can be related to ML, i.e., how these use ML.

\looseness=-1
\textbf{\textit{Business-Focused Applied}}
refers to systems that apply ML on a real-world problem and provide a benefit to end users.
They are generally focused on processing or generating data, linked with business logic to encapsulate ML.

\textbf{\textit{Conceptually Applied}}
refers to systems that demonstrate how ML technologies can be used to solve a variety of problems.
They are focused on evaluating an ML model from a data scientist's perspective and have a technical focus.

\looseness=-1
\textbf{\textit{ML Tools}}
refers to tools that support building ML models, usually libraries or frameworks that implement ML functions and algorithms, such as loss functions and optimizers (e.g. TensorFlow), but also tools that support the general ML development processes, such as experiment management tools\,\cite{idowu2021survey}.

The relation of the systems to ML is almost equally distributed, with 42\,\% of the systems using ML conceptually, 38\,\% supporting building ML models, and 35\,\% actually using ML to provide end-user-oriented functionality.

\begin{table}[t]
	\caption{Types of systems and their relation to ML}
	\vspace{-3pt}
	\label{tab:results_classification}
	\centering
	\smaller
	\begin{threeparttable}
		\begin{tabular}{l|c|cccc}
			\toprule
			                        & \textsf{total}\tnote{1} & \textsf{Application} & \textsf{Library} & \textsf{Framework} & \textsf{Plugin} \\ \midrule
			\textsf{total}\tnote{1} &                         &          32          &        99        &         32         &        2        \\ \midrule
			\textsf{Business}       &           56            &          26          &        29        &         4          &        2        \\
			\textsf{Conceptual}     &           67            &          2           &        56        &         9          &        0        \\
			\textsf{ML-Tool}        &           60            &          8           &        27        &         27         &        0        \\ \bottomrule
		\end{tabular}
		\begin{tablenotes}[para]
			\setlength{\leftskip}{0pt}
			\item [1] Total number of systems of a given type of relation to ML, independent of the other dimension. Each system can have multiple labels in each category.
		\end{tablenotes}
	\end{threeparttable}
	\vspace{-.2cm}
\end{table}

\looseness=-1
We further investigated common combinations of project type and relation to ML.
Notably, the majority of systems are libraries that apply ML conceptually (35\,\%).
Most applications (81\,\%) and all plugins are business-focused, but most business-focused systems are libraries (51\,\%).
Frameworks are mainly ML tools that allow building ML models (84\,\%).

Overall, ML projects on GitHub appear to mainly demonstrate ML functionalities and provide it to other software.

\subsection{Extent of ML-Specific Assets and Relation to Code (RQ1.2)}
\label{subsec:ml_assets}
\noindent

\begin{figure}[b]
	\vspace{-.4cm}
	\includegraphics[width=\linewidth]{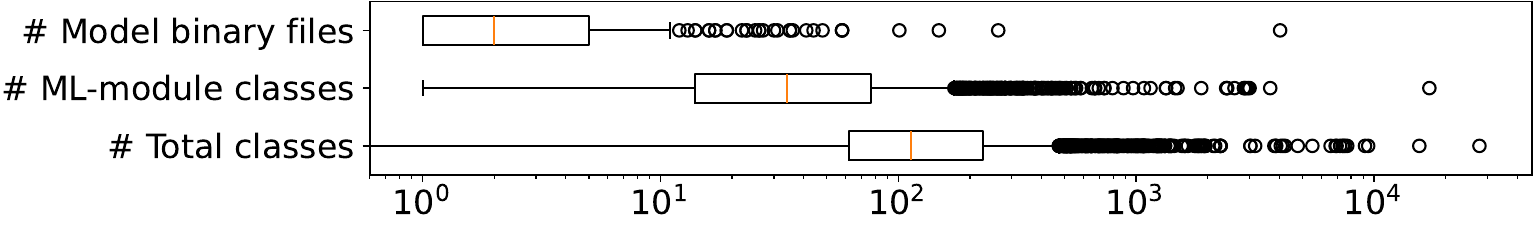}
	\vspace{-.3cm}
	\caption{Number of ML modules/ model binaries in the systems}
	\label{fig:results_models}
\end{figure}

\parhead{ML Assets}\looseness=-1\,
The most prevalent assets in our full dataset are model implementations in code, which are present in 2,499 systems (85\,\%).
The average number of classes per project that inherit from ``\textit{Module}'' or ``\textit{BaseEstimator}'' is 85, the median is 34.
These classes are either complete models or components. 
Nevertheless, it is safe to assume that the average project contains multiple ML models, since one model is unlikely to be composed of 34 modules.
Additionally, entire ML models can be loaded from binary files, containing the weights for configuring models in code.
These are, however, only contained in 339 repositories (12\,\%), with each of these containing on average 18 such files (median 2).
\Cref{fig:results_models} gives an overview of implemented modules and model binary files.
Since some forms of ML (e.g., PCA, clustering) requires no models to be stored, some systems contain no such assets.

\begin{figure}[t]
	\includegraphics[width=\linewidth]{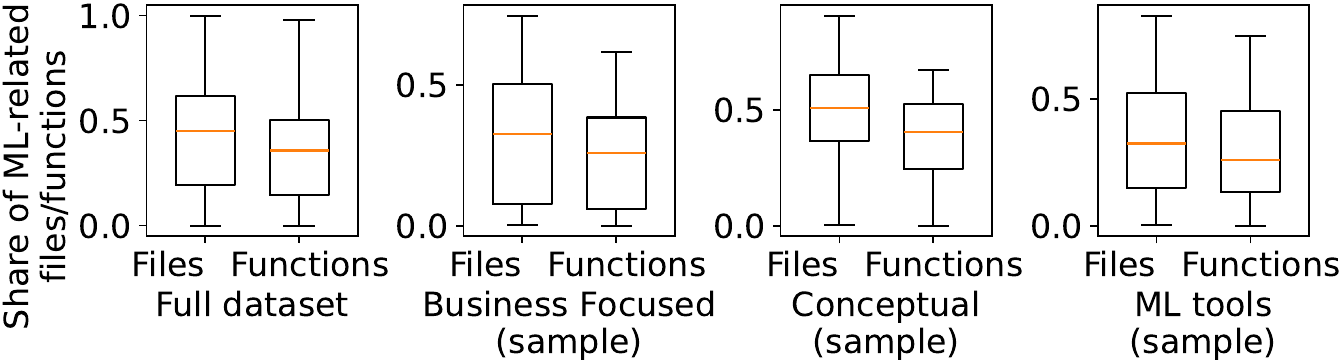}
	\vspace{-10pt}
	\caption{Share of ML code for the full dataset and the classified sample, according to the projects' relation to ML}
	\label{fig:results_ml_files}
	\vspace{-10pt}
\end{figure}

\parhead{Relation of ML and Non-ML Code}
To determine the extent of ML-related code, we measured the amount of ML-related and non-ML-related code in our full dataset.
As shown in \cref{fig:results_ml_files}, on average 42\,\% of the source files (median 45\,\%) and 34\,\% of the functions (median 36\,\%) are ML-related.
So traditional code still makes up most of the systems.

\looseness=-1
Our dataset contains systems with different relationships to ML, including those that apply ML to deliver functionality (business focused), those that demonstrate possible applications (conceptual), as well as ML tools, that support the development of other systems. We investigated on our manually classified sample how this affects the share of ML-related files and functions.
While the share of ML code is the highest for conceptual ML systems (on average 50\,\% of files and 38\,\% of functions), even in these systems, traditional source code, such as utility functions, data preprocessing, tests, and demonstrations, makes up most of the system.
In business-focused-applied ML systems, code related to ML is the lowest (on average 31\,\% of files and 24\,\% of functions) among all types of systems investigated, since ML is used only for individual features.
Similarly, ML tools provide many additional features, such as UIs, but usually focus more on ML technology. For these systems, the average share of ML code is 35\,\% of the files and 29\,\% of the functions, which is in between the other two categories.

\subsection{Testing (RQ1.3)}
\label{subsec:results_testing}
\begin{figure}[b]
	\vspace{-.2cm}
	\includegraphics[width=\linewidth]{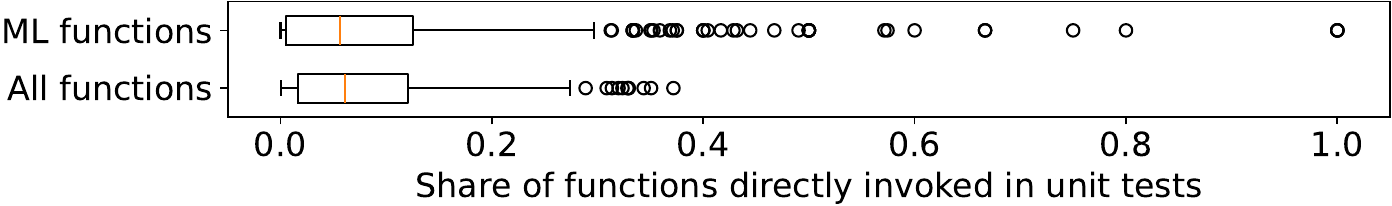}
	\vspace{-.3cm}
	\caption{Test coverage for systems with $\geq 1$ unit test}
	\label{fig:results_tests}
\end{figure}

\noindent\looseness=-1
To investigate testing practices, on the full dataset, we analyzed the prevalence of unit tests and the proportion of functions directly covered by them.
The overall prevalence of unit tests is very low, with only 954 of the subject systems (33\,\%) containing at least one unit test.
We calculated the number of functions invoked in these test cases.
Among the systems, containing at least one test file, the units tests directly cover a median of 6\,\% of all functions and 5.6\,\% of the ML-functions (see \cref{fig:results_tests}).
Test coverage of ML-enabled systems is generally rather low with no differences between ML and non-ML code.
While 90\,\% of the systems train custom ML models, 94\,\% evaluate their models.
To this end, training is implemented in a median of 5 files, and evaluation in 9 files.
Accordingly, some systems do not trust pre-trained models, but evaluate them in their application context.

\summary{Summary \textit{RQ1}: System Characteristics}{%
	The majority of ML-enabled systems on GitHub are libraries and frameworks. End-user-oriented applications are still a minority.
	The systems use ML in different ways, either developing new ML technology or applying ML on concrete problems. A small, but still substantial, number of tools that support ML engineering exist.
	The ML-enabled systems vary in size and contain many different ML assets, but also large amounts of non-ML code.
	Quality assurance is neglected.
}
\vspace{-.4cm}
\section{Reuse of ML Models and Code (RQ2)}
\label{sec:rq2}
\noindent
We now discuss the extent of model reuse in the form of pre-trained models and duplicated model implementations.

\subsection{Prevalence of Pre-Trained Models (RQ2.1)}
\label{subsec:ptm}
\begin{figure}[t]
	\centering
	\includegraphics[width=\linewidth]{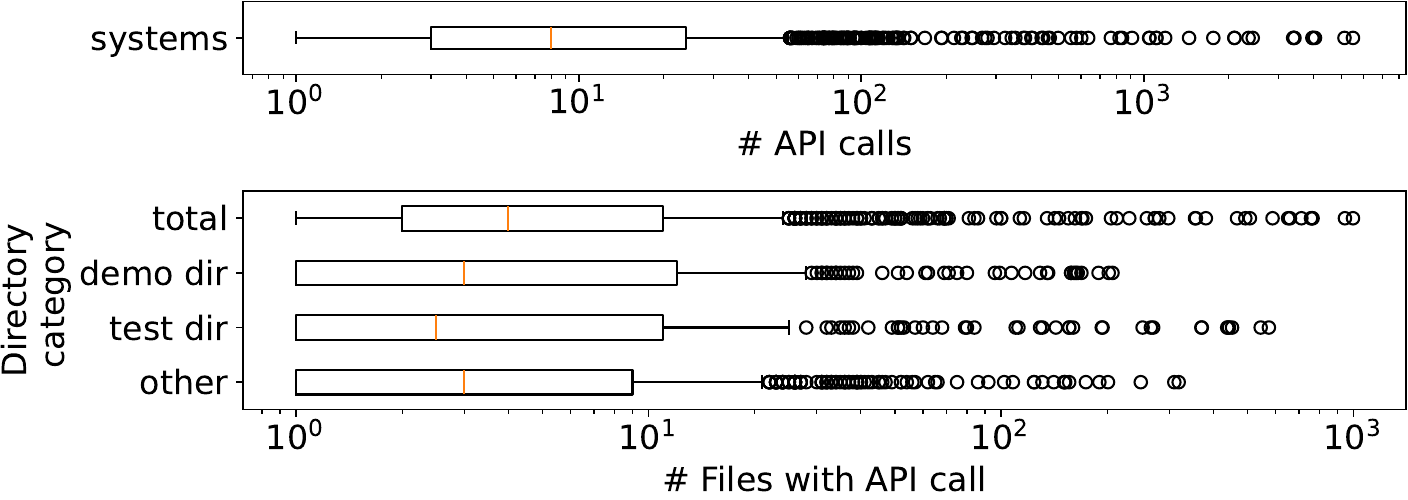}
	\vspace{-.3cm}
	\caption{Number of API calls for loading of pre-trained models per system and number of files with API call per category}
	\label{fig:results_hub_loads}
	\vspace{-.3cm}
\end{figure}

\noindent\looseness=-1
Pre-trained models, loaded from model hubs, are used in 1,209 of the 2,928 systems (41\%). The systems contain, on average, 29 API calls to the model hubs.
\Cref{fig:results_hub_loads} shows the results.

Due to the high number of API calls per system, we investigated individual systems manually, finding that often not full models are loaded but components that are used together (e.g., encoder and decoder models in Transformers or generator and discriminator in Generative Adversarial Networks).

\looseness=-1
By analyzing the directory names of code files with API calls, we determined the location in the system and potential purpose for which models are loaded.
Out of all files that contain an API call to load pre-trained models, per system, 8.3\,\% are located in test directories, and 9.8\,\% in directories named ``demo'' or ``experiment.''
The remaining 81.9\,\% are located in other directories containing code, such as model implementations, utility functions, or application logic.

\looseness=-1
ML models are loaded in 809 systems for neither testing nor demonstration purposes.
80 systems load pre-trained models only in demonstration directories.
This can be explained by the fact, that some systems in our dataset are \textit{ML Tools} (cf. \cref{subsec:system_types}), which do not implement ML-based functionality, but might demonstrate pre-trained models.

\subsection{Extent of ML Asset Duplication (RQ2.2)}
\label{subsec:code_duplicates}
\noindent
\looseness=-1
To determine how model implementations are reused, we examined for each system, how much of the source code implementing ML models is duplicated from other systems in the dataset.
We found this type of reuse in 1,690 of the subject systems (58\%).
These contain an average of 18 files that are at least partially duplicated, which is more than half of the average of 34 model implementation files in the subject systems.
We also observed that a few repositories provide the basis for most copies, and that many of these belong to the same organizations, notably Microsoft, OpenMMLab, and Hugging Face.
\Cref{tab:clusters} shows the 10 projects from which is copied the most.

\begin{table}[b]
	\centering
	\vspace{-8pt}
	\caption{Overview of the 10 most duplicated projects}
	\vspace{-0pt}
	\label{tab:clusters}
	\scriptsize
	\begin{threeparttable}
		\begin{tabular}{c}
			\begin{subtable}[t]{0.48\columnwidth}
				\centering
				\begin{tabular}{cl}
					\toprule
					\textsf{\#\tnote{1}} & \textsf{original project}  \\ \midrule
					301 & microsoft/unilm          \\
					266 & microsoft/LMOps          \\
					227 & iscyy/yoloair            \\
					193 & huggingface/transformers \\
					172 & open-mmlab/mmsegmentation \\
					\bottomrule
				\end{tabular}
			\end{subtable}
			\hfill
			\begin{subtable}[t]{0.48\columnwidth}
				\centering
				\begin{tabular}{cl}
					\toprule
					\textsf{\#\tnote{1}} & \textsf{original project}  \\ \midrule
					164 & microsoft/LoRA          \\
					145 & locuslab/convmixer      \\
					142 & open-mmlab/mmdetection  \\
					108 & facebookresearch/fairseq \\
					107 & Stability-AI/stablediffusion \\
					\bottomrule
				\end{tabular}
			\end{subtable}
		\end{tabular}
		\begin{tablenotes}[para]
			\item[1] Number of projects that copy model code from the given original.
		\end{tablenotes}
	\end{threeparttable}
\end{table}
\addtocounter{table}{-1}

\summary{Summary \textit{RQ2}: Model Reuse}{%
	Reuse of ML models is prevalent among the subject systems.
	Of the analyzed systems, 1,209 use pre-trained models and 1,690 systems copy ML implementation code.
	The copied code originates mostly from a small set of repositories maintained by an even smaller set of organizations.
}
\vspace{-.1cm}

\section{Architectural Aspects (RQ3)} 
\label{sec:rq3}
\noindent
We answer \textbf{\textit{RQ3}} based on insights from our in-depth analysis of 26 ML-enabled applications (cf. \cref{subsec:methods_manual}).

\subsection{Descriptive Aspects of ML Models}
\label{subsec:rq1_characteristics}
\noindent
We start by characterizing the role of ML models in the systems, contextualizing the results of \textbf{\textit{RQ1}} and \textbf{\textit{RQ2}}.

\parhead{ML Functionalities and Tasks}
Common functionalities provided by ML models include \textit{image processing} (e.g., \textsf{tiatoolbox}), \textit{chatbots} (e.g., \textsf{Gentopia}), \textit{video processing} (e.g., \textsf{VideoTo3dPoseAndBvh}), \textit{data analysis} (e.g., \textsf{CorpusTools}), and \textit{robotic navigation} (e.g., \textsf{sunnypilot}).
Other systems use ML to transcribe music, infer metadata, and play games.
\cref{tab:results_manual} shows the distribution of ML functions and tasks.

\looseness=-1
On a technical level, the underlying ML tasks in \cref{tab:results_manual} are \textit{generative AI} (e.g., \textsf{h2ogpt} is a chatbot), \textit{classification} (e.g. \textsf{falldetection\_openpifpaf} classifies videos), \textit{decision making} (e.g., \textsf{sunnypilot}), \textit{segmentation}  (e.g., \textsf{VideoTo3dPoseAndBvh} segments videos), \textit{dimensionality reduction} (e.g., \textsf{CorpusTools}), with 2 systems not fitting into established categories.
For example, \textsf{tournesol} uses \textit{linear regression} to visualize correlations.
Four systems combine tasks, for instance, \textsf{Video-ChatGPT} is a \textit{chatbot} with \textit{video processing} capabilities that allows user to query the content of videos.

\begin{table}[b]
	\smaller
	\setlength{\tabcolsep}{4pt}
	\vspace{-8pt}
	\caption{ML functions, tasks, and technologies (sample)}
	\label{tab:results_manual}
	\vspace{-0pt}
	\begin{threeparttable}
		\begin{tabular}{ccc}
			\begin{subtable}[t]{0.27\columnwidth}
				\centering
				\begin{tabular}{ll}
					\toprule
					\textsf{ML Function} & \textsf{\#}\tnote{1} \\ \midrule
					Image processing     & 6           \\
					Video processing     & 5           \\
					Chatbot              & 5           \\
					Data analysis        & 4           \\
					Navigation           & 3           \\
					&             \\
					&             \\ \bottomrule
				\end{tabular}
			\end{subtable}
			&
			\begin{subtable}[t]{0.33\columnwidth}
				\centering
				\begin{tabular}{ll}
					\toprule
					\textsf{ML Task}         & \textsf{\#}\tnote{1} \\ \midrule
					Generative AI            & 11          \\
					Classification           & 6           \\
					Decision Making          & 4           \\
					Segmentation             & 4           \\
					Dimensionality Reduction & 3           \\
					Clustering               & 1           \\
					&             \\ \bottomrule
				\end{tabular}
			\end{subtable}
			&
			\begin{subtable}[t]{0.36\columnwidth}
				\centering
				\begin{tabular}{ll}
					\toprule
					\textsf{ML Technology} & \textsf{\#}\tnote{1} \\
					\midrule
					Transformer   & 8  \\
					CNN           & 6  \\
					FCNN          & 5  \\
					LSTM          & 5  \\
					PCA           & 3  \\
					RL            & 3  \\
					Clustering    & 1	\\
					\bottomrule
				\end{tabular}
			\end{subtable}
		\end{tabular}
		\begin{tablenotes}[para]
			\setlength{\leftskip}{-0pt}
			\item [1] Number of systems (assignment is not exclusive, so the numbers do not add up to 26)
		\end{tablenotes}
	\end{threeparttable}
\end{table}
\addtocounter{table}{-1}

\begin{figure}[t]
	\centering
	\includegraphics[width=\columnwidth]{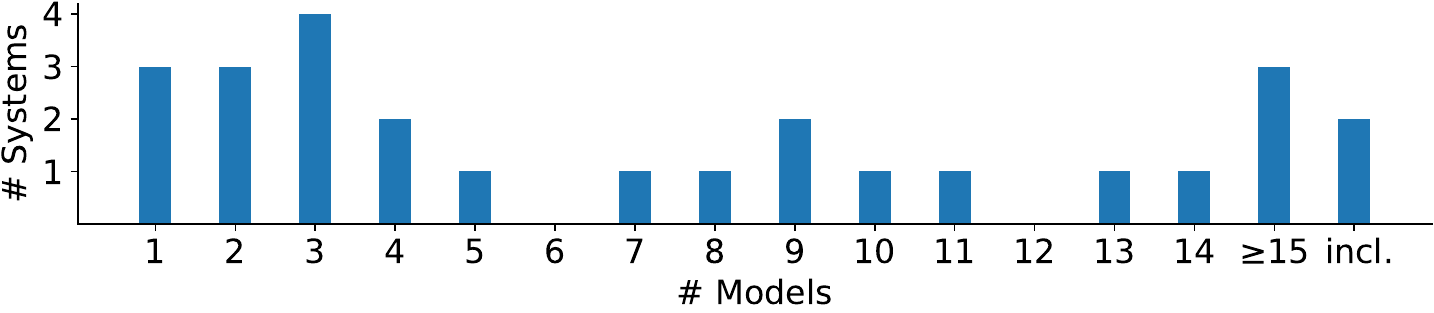}
	\vspace{-.4cm}
	\caption{Distribution of the number of models (sample)}
	\vspace{-.4cm}
	\label{fig:n_models}
\end{figure}

\looseness=-1
\parhead{Number and Type of ML Models}
The majority of the systems employ multiple ML models (see \cref{fig:n_models}), with the highest number being 18 (\textsf{h2ogpt}). Half of the systems use more than 5 models.
\looseness=-1
Three systems allow to select arbitrary models from model hubs, making the number of models practically infinite.
For example, \textsf{gentopia} allows the user to configure a \textit{chatbot} by selecting a model from Hugging Face.

\looseness=-1
The ML technologies found are shown in \cref{tab:results_manual}.
The most common type of model is a Deep Neural Network (DNN, i.e., DL) found in 21 systems.
\emph{Classical ML} (not DL) was found 5 times.
Among the DNNs, we observed multiple variations, including 8 \emph{Transformers}, 6 \emph{Convolutional Neural Networks (CNN)}, 3 \emph{Fully Connected Neural Networks (FCNN)}, and one \emph{Long Short Term Memory (LSTM)}\,\cite{hussain2022design}.
\emph{CNNs} are used exclusively for \emph{image processing}, whereas Transformers and LSTMs are mainly intended for natural language processing, although we found 3 Transformers that are used for \emph{image generation} (e.g., \textsf{ControlLoRA}).

ML can be further divided into \emph{Supervised Learning (SL)}, \emph{Unsupervised Learning (UL)}, and \emph{Reinforcement Learning (RL)} \cite{ayodele2010types}.
UL was found only in the form of classical ML, such as \emph{Principal Component Analysis (PCA)} and \emph{Clustering}.
Two systems use supervised classical ML.
The DNNs are mostly used for SL, but 3 RL systems were also found.
In contrast to DL, classical ML is mostly used for auxiliary functions, e.g., \textsf{CorpusTools} uses PCA for data visualization.

\parhead{Model Origins}
11 systems use pre-trained models directly out-of-the-box, while 7 systems fine-tune them, and 11 systems use completely custom-implemented models. 
DNNs are usually pre-trained or fine-tuned, whereas classical ML models are always custom-trained.
The likely reason is that the more complex a model becomes, the more training data and effort it requires, so pre-trained models relieve developers of end-user-oriented systems of this burden\,\cite{erhan2010does}.
We found 6 systems that use custom-trained DL models, 3 of which use RL, for which few pre-trained models exist. Other custom built ML models are used for specialized use cases, such as \textsf{DreamArtist-stable-diffusion}, which uses custom models for prompt optimization.

\looseness=-1
\parhead{Storing \& Loading of Models}
We observed local and remote storage 15 times each, with 7 systems supporting both.
For example, \textsf{Vlog} uses pre-trained models from different sources, some stored locally, some downloaded from Hugging Face.
Since UL does not need models to be stored, 3 systems have no storage.
We visualize the relationship between storage and origin of the models as a matrix in \cref{tab:storage_origin}.
Custom models are always stored locally, pre-trained models usually remotely, with some exceptions, such as \textsf{stable-diffusion-webui-depth\-map-script} or \textsf{VideoTo3dPoseAndBvh}.

\begin{table}[t]
	\centering
	\smaller
	\caption{Storage and origin of ML models (sample)}
	\vspace{-0pt}
	\label{tab:storage_origin}
	\begin{threeparttable}
		\begin{tabular}{c|l|l|ccc}
			\toprule
			& & \multicolumn{4}{c}{\textsf{Model Origin}}\\
			& & \textsf{Total}\tnote{1} & \textsf{Pre-trained } & \textsf{Fine-tuned} & \textsf{Custom} \\ \midrule
			\multirow{4}{*}{\rotatebox{90}{\shortstack{\textsf{Model}\\\textsf{Storage}}}}
			& \textsf{Total}\tnote{1}  &         & 11                    & 7                   & 11              \\ \cmidrule{2-6}
			& \textsf{Local} 	& 15             & 7                     & 4                   & 7               \\
			& \textsf{Remote}	& 15             & 9                     & 4                   & 2               \\
			&\textsf{NA}		& 3              & 0                     & 0                   & 3               \\ \bottomrule
		\end{tabular}
		\begin{tablenotes}[para]
			\setlength{\leftskip}{-4pt}
			\item[1] Number of systems with a given origin or storage, independent of the other dimension; a system can have multiple models with different origin or storage.
		\end{tablenotes}
	\end{threeparttable}
	\vspace{-8pt}
\end{table}

\subsection{Model Embedding (RQ3.1)}
\noindent
ML models are embedded in traditional code by using code-level APIs.
The main interaction with models is via their inputs and outputs, which often require pre- or post\-processing.
However, we observed that developers often resort to ad hoc solutions to embed ML models.
We report on observed inputs and outputs, and pre- and postprocessing practices.

\parhead{Model In- \& Output}
We found a wide range of data types used as inputs and outputs to ML models (see \cref{tab:input_output}).
Inputs include text (e.g., \textsf{paperless-ng}), images (e.g., \textsf{imagepy}), and video (e.g., \textsf{VideoTo3dPoseAndBvh}).
Less frequent are audio (\textsf{sheetsage}) or an environment state for RL agents, in the form of structured numeric data (\textsf{deep\_learning\_and\_the\_game\_of\_go} or \textsf{sarl\_star}).
The outputs are more diverse, commonly text (e.g., \textsf{sheetsage)}, images (e.g., \textsf{clip-glass}), labels (e.g., \textsf{paperless-ng}, which classifies text documents), actions (e.g., \textsf{sunnypilot}, or other robotics systems), bounding boxes (e.g., \textsf{VIAME}), and video, (e.g., \textsf{VLog}).
While we mostly observed standard data formats, special application scenarios sometimes require less common formats, e.g., \textsf{VideoTo3dPoseAndBvh} extracts animations from video and saves in the BVH format.
Structured numeric data is outputted by the model of \textsf{Sunnypilot} containing inferred information about the environment, like the lane lines or the distance to the lead vehicle.
The most common combination of input and output is text to text (4 systems, all chatbots).
Other common combinations are text to image or image to text. \textsf{clip-glass} is an example of both, providing image captioning and image generation from text.

\begin{table}[b]
	\vspace{-8pt}
	\setlength{\tabcolsep}{4pt} 
	\renewcommand{\arraystretch}{1.2} 
	\smaller
	\centering
	\caption{Data types as input and output to models (sample)}
	\vspace{-3pt}
	\label{tab:input_output}
	\begin{tabular}{lcccccccc}
		\toprule
		& \textsf{Text}
		& \textsf{Image}
		& \textsf{Video}
		& \textsf{Labels}
		& \textsf{Actions}
		& \parbox{.95cm}{\centering \textsf{Numeric}\\ \textsf{Data}}
		& \parbox{.95cm}{\centering \textsf{Bounding}\\ \textsf{Boxes}}
		& \textsf{Audio} \\
		\midrule
		Input  & 10   & 10    & 5     & -     & -      & 2       & -                 & 1    \\
		Output & 8    & 5     & 2     & 4     & 4      & -       & 2                 & -	\\
		\bottomrule
	\end{tabular}
\end{table}

\parhead{Pre- \& Postprocessing}
Almost all systems contain pre- and post processing steps, ranging from generic operations (e.g., text tokenization) to more complex content manipulations, often tailored to the needs of a specific system.
We clearly identified preprocessing code in 15 systems; 4 systems contained no preprocessing; while the remaining 7 were inconclusive.
Common preprocessing steps include normalization, feature extraction, and vectorization.
Some processing functions such as image normalizations were scattered throughout the system.

Postprocessing was found in 8 systems, 9 contain none, 9 were inconclusive.
Again, the majority are simple data conversions, such as projecting tokens back to text.
More complex tasks were also observed, e.g., validating actions chosen by an RL agent (\textsf{deep\_learning\_and\_the\_game\_of\_go}).

Particularly interesting is \textsf{tiatoolbox}, which provides users with extensive configuration options, including pre- and post-processing methods.
While it provides default functions, it also supports the integration of user-written functions added to the model as callbacks.
\textsf{DreamArtist-stable-diffusion}, implements some security functionalities that ensure input integrity.

\begin{figure}[t]
	\centering
	\includegraphics[width=\linewidth]{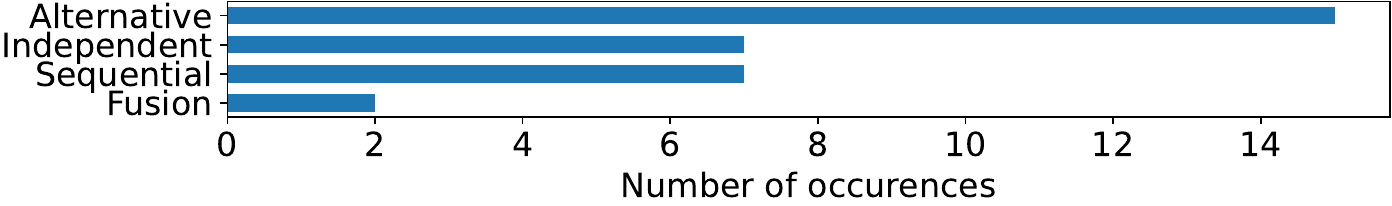}
	\vspace{-.4cm}
	\caption{Distribution of model interaction patterns (sample)}
	\label{fig:model_interactions}
	\vspace{-.4cm}
\end{figure}

\subsection{Interaction Patterns (RQ3.2)}
\label{subsec:interaction_patterns}
\noindent
In the 23 systems with multiple models (see \cref{sec:rq1}), we observed various types of interaction between these models.
Triangulating with related work\,\cite{peng2020apollo,Peldszus2023}, we extracted 4 interaction patterns that describe the observed topologies. \Cref{fig:model_interactions} shows the frequency of each pattern in the sample.

\textbf{\textit{Alternative}} use of models is the most common pattern (15 systems), where multiple models are offered for the same task and chosen either automatically or by the user.
For example, \textsf{sarl\_star }provides multiple RL agents to control a robot.

\textbf{\textit{Independent}} models for unrelated functionalities are found in 7 systems.
Larger systems in particular provide the user with multiple functions, some of which require ML.
For example, \textsf{VIAME} offers \textit{image processing} tools (e.g., for \textit{segmentation} or \textit{classification}) to be used independently by the user.

\textbf{\textit{Sequential}} integration of ML models was observed in 7 systems.
The output of one model is used as input to another model, possibly with additional processing steps in between.
This pattern is used for more complex tasks, such as perception systems; e.g., \textsf{home-robot} uses a model to processes sensor input and feeds the results (i.e., detected obstacles) to an RL agent for navigation.
Sometimes, additional data is added to the subsequent models, e.g, \textsf{Video-ChatGPT} uses a Transformer model to summarize videos, which is the input for a Large Language Model (LLM) together with text~prompts.

\textbf{\textit{Joining}} integration of ML models is the most complex pattern and was found twice in different variants. Results of two or more models are combined, either by another model (which also includes the sequential pattern) or code logic.
We observed one variant in \textsf{Vlog}, which has a complex pipeline of 6 models to transform videos into text descriptions.
It includes a model transcribing audio and one generating captions for each frame. Both results are fed into an LLM, which assembles a final text document.
The other variant was observed in \textsf{eynollah}, where two segmentation models process an image, with each model being used for different parts of the image. The segmented image is further processed by other models.

Multiple of these patterns are implemented in 8 systems.
For instance, \textsf{VIAME} offers multiple \textit{independent} ML functionalities, but has \textit{alternative} models for some of them.
In \textsf{Vlog}, data processing starts with a feature extractor model, followed by a segmentor (\emph{sequential}) to divide a video into chunks.
The chunks are then processed by two models, one for subtitle generation and the other for segmentation and classification.
In parallel, the audio track is translated into text by a third ML model.
The outputs of the three models are then assembled by an LLM (\emph{joining}) into a text description of the video.

\summary{Summary \textit{RQ3}: Architectural Aspects}{
	Most systems use supervised DL with pre-trained models, followed by deep reinforcement learning and classical unsupervised learning. While most ML models process text or images, other types of input are also present. Using multiple models is very common. Although most models are either used as alternatives or independently, some systems apply multiple models sequentially.
	Pre-and postprocessing steps are scattered throughout the codebase and there are no common practices for integrating ML models.
}
\vspace{-.1cm}
\section{Related Work}
\label{sec:relwork}
\noindent\looseness=-1
Prior research has characterized ML-enabled systems and examined challenges related to the integration of ML and relevant aspects.
We complement with the first investigation of model reuse and architectural aspects on a large dataset.

\looseness=-1
Similar to how we study \textbf{system characteristics}, Gonzales et al.\,\cite{gonzalez2020state} analyze ML projects, compare them to non-ML projects, and identify unique properties, such as programming languages, library dependencies, and developer collaboration.
We confirm some of their findings, for instance, that ML tools are still more popular than applied ML projects.
Our study, however, has a much wider scope, including a qualitative analysis that we relate to the quantitative analysis.

\looseness=-1
Jiang et al.\,\cite{jiang2024ptm} study \textbf{model reuse} on a dataset of pre-trained models.
Their dataset is limited, considering only two major model hubs, whereas we consider seven. We also investigate reuse of ML assets through code duplication.

\looseness=-1
Gorton et al.\,\cite{gorton2023software} suggest a conceptual \textbf{ML architecture}, which we did not observe in the analyzed systems.
Peng et al.\,\cite{peng2020apollo} examine the architecture of an autonomous driving system and identify patterns of interaction between ML models and source code.
We complement their findings with a broader perspective, covering 26 ML-enabled systems.
Similar to us, Nahar et al.\,\cite{nahar2024product} combine repository mining with a manual analysis of a subset.
They find that only a fraction of systems using multiple ML models contain interacting models.
However, their analysis is at a more abstract level than ours, focusing on processes and collaboration rather than implementation details such as the type of ML model used. They also focus on qualitative methods, while we combine these with quantitative analysis.

Houerbi et al.\,\cite{houerbi2024empirical} studied \textbf{CI pipelines} in open source ML projects and identified significant issues.
In fact, in our dataset, 1,091 of 2,928 repositories (37\%) use continuous integration (883 use GitHub actions, 192 Travis, and 16 Jenkins), confirming their observations about adoption.
Openja et al.\,\cite{openja2024empirical} investigate \textbf{testing practices} by manually analyzing test files from 11 open-source ML projects, highlighting used test types and tested ML-related functionalities.

\looseness=-1
While we took a static view on ML-enabled systems, others analyze their \textbf{evolution}, focusing on how ML assets evolve in relation to source code\,\cite{barrak2021co}, and the types of changes contributed by forks of ML repositories\,\cite{bhatia2022towards}. Simmons et al.\,\cite{simmons2020large} analyze issues in open-source ML projects, finding that ML-related issues take longer to resolve than non-ML issues.

Finally, Munappy et al.\,\cite{munappy2022data} investigate challenges in \textbf{data management} for deep learning models, identifying 20 challenges and possible solutions.
Biswas et al.\,\cite{biswas2021art} characterize data-science pipelines,
finding that in practice they differ from each other and are often more complicated than the theoretical models.
This is, however, beyond the scope of our study.
\section{Discussion}
\label{sec:discussion}
\noindent
We now discuss our findings and formulate actionable recommendations for researchers and practitioners.

\looseness=-1
\parhead{ML Adoption in Open-Source Software}
Despite growing attention, the adoption of ML in open-source systems is lower than expected.
Most systems are prototypes (e.g., \textsf{ConsistentTeacher}, a research prototype to improve object detection)
and proofs of concept, demonstrating new ML technologies (e.g., \textsf{disrupting-deepfakes}, which demonstrates how images can be protected from manipulation).
Applications providing end-user-oriented functionality (e.g., \textsf{paperless-ng}, a tool for organizing scanned text documents by inferring metadata) are a minority and often appear prototypical as well (e.g., \textsf{home-robot} refers to itself as a research tool).
This aligns with others' findings\,\cite{nahar2024product}, which suggest that open-source ML applications often resemble startup-style projects where ML is not a central component.

While the analyzed applications offer diverse capabilities based on ML, these are often based on ad hoc solutions to integrate ML models (e.g., \textsf{falldetection\_openpifpaf} uses 19 if-statements to load alternative models).
This indicates that developers are still exploring the possibilities of ML (cf. \textbf{\textit{RQ1.1}}), and highlights the need for design methods that consider the specifics of ML and more structured techniques to integrate ML models into software systems (cf. \textbf{\textit{RQ3.1}}).

\summary{Recommendation}{%
	Effective ML adoption requires a holistic perspective on building ML-enabled systems. Further studies are needed to understand ML adoption barriers (e.g., lack of use cases, engineering practices, or resources), and develop methods and tools for adoption.
}

\looseness=-1
\parhead{ML Asset Reuse}
ML assets are extensively reused across software systems, often employing bad reuse practices, such as code cloning.
We identified reuse as an essential practice, as 65\,\% of systems reuse ML assets.
This reuse can be easily motivated, e.g., due to lack of data science expertise among software engineers\,\cite{hopkins2021developmentresources}.
However, the observed reuse is rarely systematic, i.e., cloning model code, which poses maintainability issues---as observed in code clones in general\,\cite{kapser2008cloning,Juergens2009}.
Although pre-trained models should provide a manageable form of reuse, 75\,\% of the systems using pre-trained models contain also code clones.
While we confirm the importance of pre-trained models in practice, other studies also highlight issues, such as insufficient security measures, missing attributes or artifacts, and discrepancies in model performance\,\cite{jiang2023huggingface,jiang2022reusehub}.
We show that code duplication is not predominant in a single ML technology, but prone to all areas, i.e., transformers, stable diffusion, and others (cf. \cref{tab:clusters}).
Our in-depth analysis of a sample of 26 ML-enabled applications, revealed no other patterns of model reuse.
A possible reason for cloning model implementations is that developers modify the model implementations slightly, which is supported by the fact, that many of the identified code duplicates are not complete matches.
However, this may also result from duplicates using outdated model versions.

\summary{Recommendation}{%
	Increase developers' awareness of proper reuse mechanisms, particularly for ML assets, and promote their adoption.
	This must be accompanied by academic studies to identify the reasons for the observed bad practices and the development of effective, easy-to-use reuse mechanisms.
}

\looseness=-1
\parhead{Topologies of ML Models}
\textit{Many ML-enabled systems integrate multiple models. However, the actual embedding of ML models into the systems and the integration of multiple models is realized in an ad hoc manner. Best practices are missing.}
In line with other studies\,\cite{peng2020apollo,nahar2024product}, the analyzed systems often employ multiple ML models that interact in various ways through complex code logic.
For example, Nahar et al.\,\cite{nahar2024product} show that 60\,\% of th systems they investigated have multiple models and 23\,\% contain interacting models.
In our sample, even 88\,\% of the systems contain multiple models, and 69\,\% involve some form of interaction.
Still, we did not observe any systematic integration strategy, but ad hoc implementations. Especially pre- and post processing of data lacks clear structure (e.g., \textsf{tiatoolbox} allows for arbitrary pre-and postprocessing methods, other systems, e.g., \textsf{Gentopia} contain none).
Other observations include heavy use of wrappers for loading ML models, and extensive IF-statements for loading alternative models (\textsf{falldetection\_openpifpaf} contains an IF with 19 alternatives).
This lack of methods, missing architectures or templates, may be a core reason why advancements in ML do not carry over to building systems (in fact, most projects stall or fail in the development phase\,\cite{ONeill2019}).
Researchers can assist by providing guidelines on best practices and architectures for developing complex topologies of ML models.
Additionally, tools and libraries, such as model hubs, should be improved to provide adequate support for developing ML-enabled systems, particularly considering multiple ML models.
Current tools, such as experiment management tools, are not only model-centric, but lack support for multiple models\,\cite{idowu2021survey,idowu2022effectiveness}.

\summary{Recommendation}{%
	Complex ML model topologies are part of ML-enabled systems and need to be considered in future research.
	This requires design patterns and tools that can manage multiple models, lifting the current model-centric to a system-centric perspective in the development of ML-enabled~systems.
}

\section{Conclusion}
\label{sec:conclusion}

\noindent
We presented a large-scale study of ML model integration in ML-enabled software systems.
We quantitatively analyzed 2,928 systems in terms of their characteristics, comparing ML and non-ML assets and reuse practices.
We manually classified a random sample of 160 systems in terms of their relationship to ML and system type, and analyzed the 26 applications of this sample in depth regarding their provided functionality, model reuse practices, and architectural aspects.

Among others, we find that ML is used for a variety of functionalities, but mostly in conceptual prototypes.
We discovered extensive, but unstructured reuse of ML models.
Although we observed a lack of structured approaches to the architecture of ML-enabled systems, we identified four patterns of interaction between ML models.
We contribute recommendations for researchers and tool builders.

\section*{Acknowledgment}
\noindent
We thank Kevin Hermann for providing feedback and Constanze Ohrem for the help with labeling the dataset. This work was partially funded by the German Federal Ministry for Education and Research (BMBF) under the project PrivacyE2E.
\balance

\bibliographystyle{ieeetr}
\bibliography{doc}

\end{document}